\documentclass[aps,prb,superscriptaddress,showpacs,twocolumn]{revtex4}

\usepackage{color}
\usepackage{amssymb}
\usepackage{graphicx}
\usepackage{amsmath}
\usepackage{dcolumn}
\usepackage{bm}
\begin{document}

\title{Enhanced Spin-Flip Scattering by Surface Roughness in WS$_2$ and MoS$_2$ Armchair Nanoribbons}
\author{Shoeib Babaee Touski}
\affiliation{Department of Electrical Engineering, Hamedan University of Technology, Hamedan 65155, Iran}
\author{Rafael Rold\'an}
\email{rroldan@icmm.csic.es}
\affiliation{Instituto de Ciencia de Materiales de Madrid, CSIC, E-28049 Cantoblanco, Madrid, Spain}
\author{Mahdi Pourfath}
\email{pourfath@ut.ac.ir}
\affiliation{School of Electrical and Computer Engineering, University of Tehran, Tehran 14395-515, Iran}
\affiliation{School of Physics, Institute for Research in Fundamental Sciences (IPM), Tehran 19395-5531, Iran}
\affiliation{Institute for Microelectronics, Technische Universit\"at Wien, Gu{\ss}hausstra{\ss}e 27--29/E360, A-1040 Wien, Austria}
\author{M. Pilar L\'opez-Sancho}
\affiliation{Instituto de Ciencia de Materiales de Madrid, CSIC, E-28049 Cantoblanco, Madrid, Spain}

\date{\today}

\begin{abstract}
The band structures of single-layer  MoS$_2$ and WS$_2$  present a coupling between spin and valley degrees of freedom that suppresses spin-flip scattering and spin dephasing.
Here we show that out-of-plane deformations, such as corrugations or ripples,  enhance spin-flip scattering in armchair MoS$_2$ and WS$_2$ nanoribbons.
Spin transport in the presence of surface roughness is systematically investigated, employing the non-equilibrium Green's function method along with the tight-binding approximation. Both transmission and conductance have been calculated as a function of surface roughness. Our results indicate that the spin-flip rate, usually neglected in flat pristine samples,  increases significantly with the surface roughness amplitude.
These results are important for the  design and fabrication of transition metal dichalcogenides based spintronic devices.

\end{abstract} 


\maketitle
\section{Introduction}
\label{intro}
A long spin relaxation length, i.e. the possibility for spin-polarized electrons to travel larger distances without losing encoded information, is a basic requirement for spintronic applications.  Graphene was envisioned early on as a promising material for spintronics, owing to the combination of the unique electronic band structure of so-called massless Dirac fermions, weakly sensitive to backscattering and traveling at very high speed over very large distances, even at room temperature.~\cite{Han14B,Roche15} However the weak spin-orbit coupling (SOC) in graphene makes spin effects very small.  The discovery of graphene  paved the way for investigating other two-dimensional (2D) materials with properties complementary to those of graphene.\cite{novoselov12} Stacking of different families of 2D materials in a controlled fashion can create heterostructures with tailored properties that offers promising avenues to design and fabricate novel devices.\cite{Geim13}

Single layers of transition metal dichalcogenides (TMDC) like MoS$_2$ and WS$_2$ are direct band gap semiconductors with strong spin-orbit coupling, which originates from $d$-orbitals of 
the heavy transition metal atoms. This allows for the control of spin with electric field.\cite{Klinovaja_PRB13,Ochoa_PRB13,Zibouche_PRB14,Kormanyos_PRX14,kosmider2013electronic,georgiou2013vertical} The band structure of TMDC consists of two inequivalent valleys (K and K') located at the corners of the hexagonal Brillouin zone.\cite{lebegue09elec} The lack of inversion symmetry alongside the large SOC strength lead to the coupling of spin and valley degrees of freedom,\cite{zeng2013optical} allowing for spin and valley control with the potential use of TMDCs in valleytronics and spintronics.\cite{zeng2012valley,sallen2012robust,xiao2012coupled,Song_PRL13} Magneto-transport experiments have estimated an upper limit for the spin-orbit scattering length for $n$-type MoS$_2$ as high as 430~nm.\cite{neal13magneto} However mirror symmetry along the $z$-direction is usually broken due to surface ripples, thermal out-of-plane fluctuations, sulfur vacancies, etc., what leads to a Rashba-like spin-orbit contribution which can limit the spin lifetime.\cite{brivio11ripples} In particular, static wrinkles have been shown to affect more the spin coherence as compared to out-of-plane phonons.\cite{ochoa2013spin} 

\begin{figure}[t]
    \includegraphics[width=1.\linewidth]{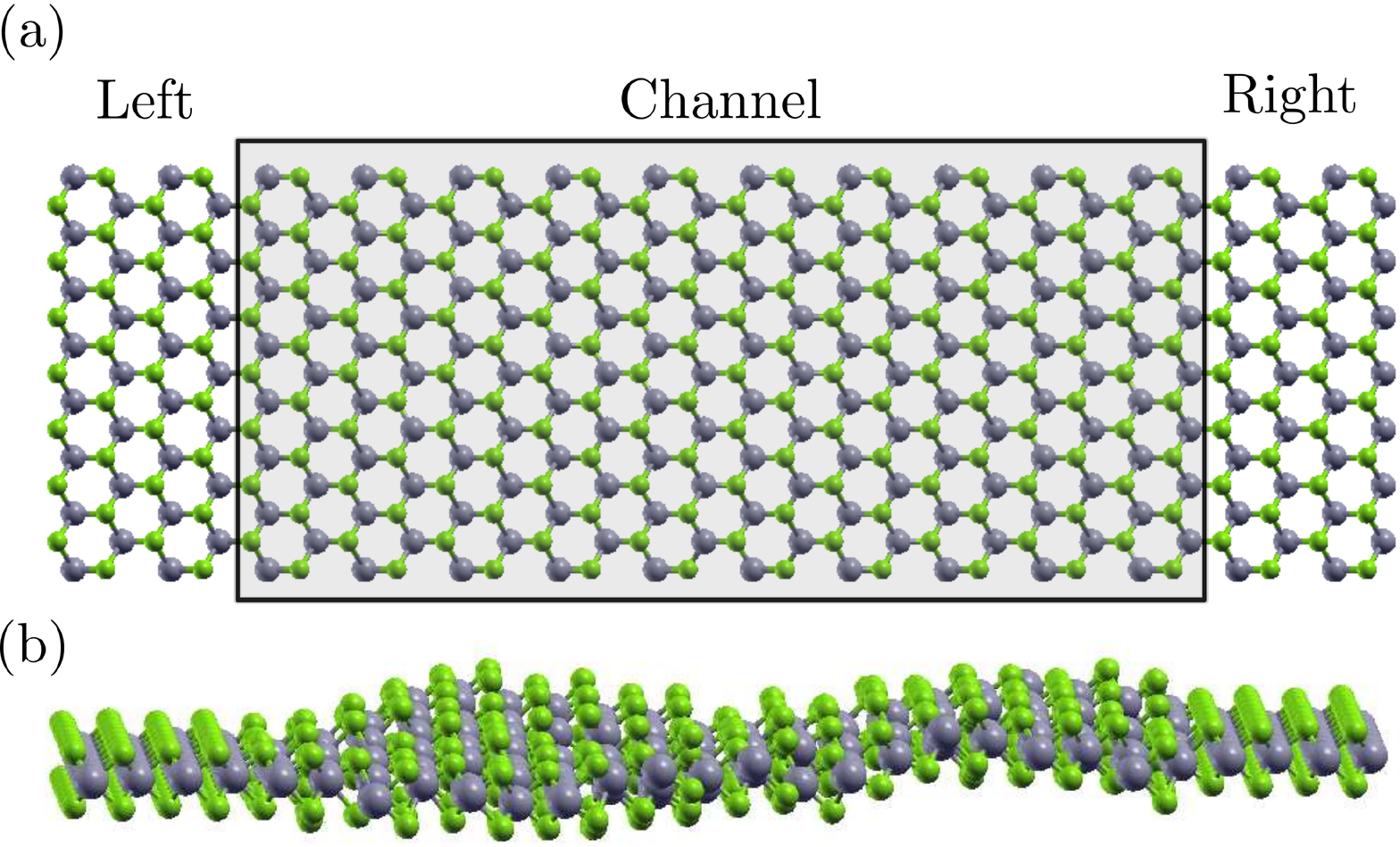}
  \caption{ Sketch of the system considered in our calculations. (a) Channel and leads are made of the same family of TMDC, MoS$_2$ or WS$_2$. (b) Surface roughness is considered for the channel sector, but not for the leads. }
  \label{f:Sketch}
\end{figure}

Nanoribbons of TMDCs can be obtained by tailoring a ribbon from an infinitely extended mono-layer,\cite{liu13} and can be synthesized by using electrochemical methods,\cite{li04poly} or by unzipping nanotubes.\cite{wang10mixed,nethravathi13,vasu15} First-principle calculations indicate that armchair MoS$_2$ and WS$_2$ nanoribbons show semiconducting behaviour, while zigzag nanoribbons are metallic.\cite{li08mos2,lopez2015electronic} In this work we study the effect of surface corrugation on spin-transport in armchair nanoribbons of MoS$_2$ and WS$_2$ (Fig. \ref{f:Sketch}). For this we use  non-equilibrium Green's function formalism along with a tight-binding model applied to nanoribbons of TMDCs in the presence of surface roughness. We find a significant increase of spin-flip rate due to static corrugations of the sample. The effects are more important in WS$_2$ than in MoS$_2$, due to the stronger atomic SOC of W atoms. Our results suggest that the use of flat substrates can considerably increase the efficiency of TMDCs for spintronics applications.  

The paper is organized as follows. In Sec. \ref{Sec:Approach}  we describe the model and the calculation method, taking into account  surface rougness effects. In Sec. \ref{Result} the results are presented and discussed. Finally, in Sec. \ref{Sec:Conclusions} the main conclusions are summarized. In Appendix \ref{App:Periodic} we present results obtained considering periodic boundary conditions.

\section{Model and Method}\label{Sec:Approach}
\subsection{Thight Binding Hamiltonian}\label{Sec:TB}

The crystal structure of TMDCs consist of one triangular lattice of metal atoms (Mo or W in the present case) which is sandwiched between two triangular lattices of chalcogen (S) atoms. Thus the unit-cell contains one transition-metal atom plus two chalcogen atoms. Our calculations will be performed by using a tight-binding model that contains five $d$ orbitals of the metal (Mo/W) atoms and three $p$ orbitals for each of the two calchogen S atoms in the unit cell.\cite{cappelluti13} Therefore the dimension of the Hamiltonian of a single layer  (before including spin degrees of freedom) will be $11\times11$, and can be written as:
\begin{equation}
\hat{H_0}=\sum_{i;l} \epsilon_{i;l} \hat{c}_{i;l}^\dagger \hat{c}_{i;l} + \sum_{\left\langle i,j \right\rangle;l,m} V_{i,j;l,m}\hat{c}_{i;l}^\dagger \hat{c}_{j;m}
\end{equation}
where $i,j$ are the atomic position indices, $l$ and $m$ label the atomic orbitals, $\hat{c}_{i;l}^\dagger (\hat{c}_{i;l})$ creates (annihilates) an electron at orbtial $l$ of site $i$, $\epsilon_{i;l}$ refers to on-site crystal fields of orbital $l$ and $V_{i,j;l,m}$ are hopping parameters, where $\langle ij\rangle$ runs over first nearest neighbor sites. The Slater-Koster parameters for MoS$_2$ and WS$_2$ obtained from fits to density functional theory (DFT) calculations are reported in Table \ref{Tab:Parameters}.\cite{Rostami2015b}  The ratio of the metal $d$ and chalcogen $p$ orbital contributions to the valence and conduction bands in our tight-binding model is $\sim70/30$ respectively, in good agreement with DFT calculations.\cite{cappelluti13} By performing a unitary transformation that accounts for the symmetric and antisymmetric combinations of S-$p$ orbitals of the top and bottom chalcogen atoms, it is possible to simplify the model into two decoupled blocks of dimensions $6\times6$ and $5\times5$, being the $6\times6$ block the relevant subspace for low energy calculations since it contains the valence and conduction band states. The bases of these blocks are $(d_{xy},d_{x^2-y^2},d_{3z^2-r^2},p_x^S,p_y^S,p_z^A)$ and $(d_{xz},d_{yz},p_x^A,p_y^A,p_z^S)$, respectively,\cite{cappelluti13} where $A$ and $S$ stand for the symmetric and antisymmetric combinations of the top $t$ and bottom $b$ chalcogen $p$ orbitals, $p_i^{S(A)}=1/\sqrt{2}(p_i^t\pm p_i^b)$, where $i=x,y,z$.

\begin{table}[t]
\begin{tabular}{lclcrcr}
\hline
\hline
                          &                                &                   &                                 &    MoS$_2$     & \hspace{0.1truecm} & WS$_2$ \\
\hline
\\
  SOC   & \hspace{0.1truecm} &$\lambda_{M}$ & \hspace{0.1truecm} &  0.075 & \hspace{0.1truecm} & 0.215 \\
                        & \hspace{0.1truecm} &$\lambda_{\rm S}$ & \hspace{0.1truecm} &  0.052 & \hspace{0.1truecm} & 0.057 \\
\\
   Crystal Fields & \hspace{0.1truecm} &$\epsilon_{d_{3z^2-r^2}}$  & \hspace{0.1truecm} &  -1.094 & \hspace{0.1truecm} & -0.872 \\
                          & \hspace{0.1truecm} &$\epsilon_{d_{xz}},\epsilon_{d_{yz}}      $ & \hspace{0.1truecm} &  0.670         & \hspace{0.1truecm} & 0.670 \\
                          & \hspace{0.1truecm} &$\epsilon_{d_{xy}},\epsilon_{d_{x^2-y^2}} $ & \hspace{0.1truecm} & -1.511  & \hspace{0.1truecm} & -1.511  \\
                          & \hspace{0.1truecm} &$\epsilon_{p_x},\epsilon_{p_y}$  & \hspace{0.1truecm} &  -3.559 & \hspace{0.1truecm} & -3.468 \\
                          & \hspace{0.1truecm} &$\epsilon_{p_z}$ & \hspace{0.1truecm} &  -6.886 & \hspace{0.1truecm} & -3.913  \\
\\                
$M$-S & \hspace{0.1truecm} &$V_{pd\sigma}$ & \hspace{0.1truecm} &  3.689 & \hspace{0.1truecm} & 3.603 \\
                          & \hspace{0.1truecm} &$V_{pd\pi}$ & \hspace{0.1truecm} &  -1.241  & \hspace{0.1truecm} & -0.942 \\
\\                      
$M$-$M$& \hspace{0.1truecm} &$V_{dd\sigma}$ & \hspace{0.1truecm} &  -0.895 & \hspace{0.1truecm} &  -1.216\\
                          & \hspace{0.1truecm} &$V_{dd\pi}$ & \hspace{0.1truecm} &  0.252 & \hspace{0.1truecm} &  0.177\\
                          & \hspace{0.1truecm} &$V_{dd\delta}$ & \hspace{0.1truecm} &  0.228 & \hspace{0.1truecm} &  0.243\\
\\                        
S-S & \hspace{0.1truecm} &$V_{pp\sigma}$ & \hspace{0.1truecm} &  1.225 & \hspace{0.1truecm} &  0.749\\
                          & \hspace{0.1truecm} &$V_{pp\pi}$ & \hspace{0.1truecm} &  -0.467 & \hspace{0.1truecm} &  0.236 \\                        
\hline
\hline
\end{tabular}
\caption{Spin-orbit coupling $\lambda_{\alpha}$ and tight-binding parameters for single-layer $M$S$_2$, where the metal $M$ is Mo or W. All the Slater-Koster parameters are in units of eV.}
\label{Tab:Parameters}
\end{table}

Spin-orbit coupling, however, mixed these blocks through processes that flip the electron spin.\cite{Roldan14mom} The SOC contribution is included in our theory trough the term
\begin{equation}
\hat{H}_\mathrm{SO}= \sum_{i;l,m}\frac{\lambda_{i;l,m}}{\hbar} \hat{L}_{i;l}\cdot \hat{S}_{i;m},
\end{equation}
where $\lambda$ is the intra-atomic SOC constant, $\hat{L}$ is the angular momentum operator for atomic orbitals, and $\hat{S}$ is the spin operator.  It is useful to express $\hat{H}_\mathrm{SO}$ as:
\begin{equation}\label{Eq:HSO}
\hat{H}_\mathrm{SO}= \sum_{i;l,m}\frac{\lambda_{i;l}}{\hbar} \left[ \frac{\hat{L}_{i;l}^+ \hat{S}_{i;m}^- + \hat{L}_{i;l}^- \hat{S}_{i;m}^+}{2} + \hat{L}_{i;l}^z \hat{S}_{i;m}^z \right]
\end{equation}
where ${\hat{\cal O}}^{\pm}={\hat{\cal O}}^{x}\pm i{\hat{\cal O}}^{y}$ are the standard ladder operators, with ${\hat{\cal O}}={\hat L},{\hat S}$. We can distinguish two different contributions to the SOC Hamiltonian (\ref{Eq:HSO}), the first term which leads to spin-flip processes, and the spin-conserving term $\propto \lambda \hat{L}^z\hat{S}^z$. For flat pristine MoS$_2$ or WS$_2$, spin-flip processes are negligible and full spin polarization as well as long spin relaxation lengths can be achieved.\cite{Xu_NP14} In this limit one can safely reduce to the $6\times6$ block.\cite{Roldan14mom} However, as experimentally observed \cite{chun14,wencan15} realistic samples do not preserve mirror symmetry along the $z$-direction. This is due to the presence of sulfur vacancies, or to corrugations and ripples in the sample, associated e.g. to the presence of a substrate or due to thermal out-of-plane phonons. In this situation, the $6\times 6$ and the $5\times 5$ blocks are coupled. As a consequence, the  contribution of $d_{xz}$ and $d_{yz}$ orbitals to the density of states (DOS) of the corrugated ribbon is  significantly larger than in the flat situation. Importantly, spin-flip processes become relevant, limiting spin life time.\cite{brivio11ripples}  Therefore we will use in our calculations the whole Hilbert space of dimension $2\times11$ (including spin). We notice that, contrary to previous works that consider the effect of flexural phonons, corrugations or topological defects in the transport properties from minimal ${\bf k}\cdot{\bf p}$ models and group theory methods,\cite{Song_PRL13,Ochoa2016} here we use a tight-binding model that accurately accounts for the states of the valence and conduction bands in the whole Brillouin zone, to calculate spin-resolved transmission probabilities through a finite corrugated armchair ribbon of MoS$_2$ or WS$_2$.

\subsection{Non-equilibrium Green's function method}

\begin{figure*}[t]
    \includegraphics[width=0.8\linewidth]{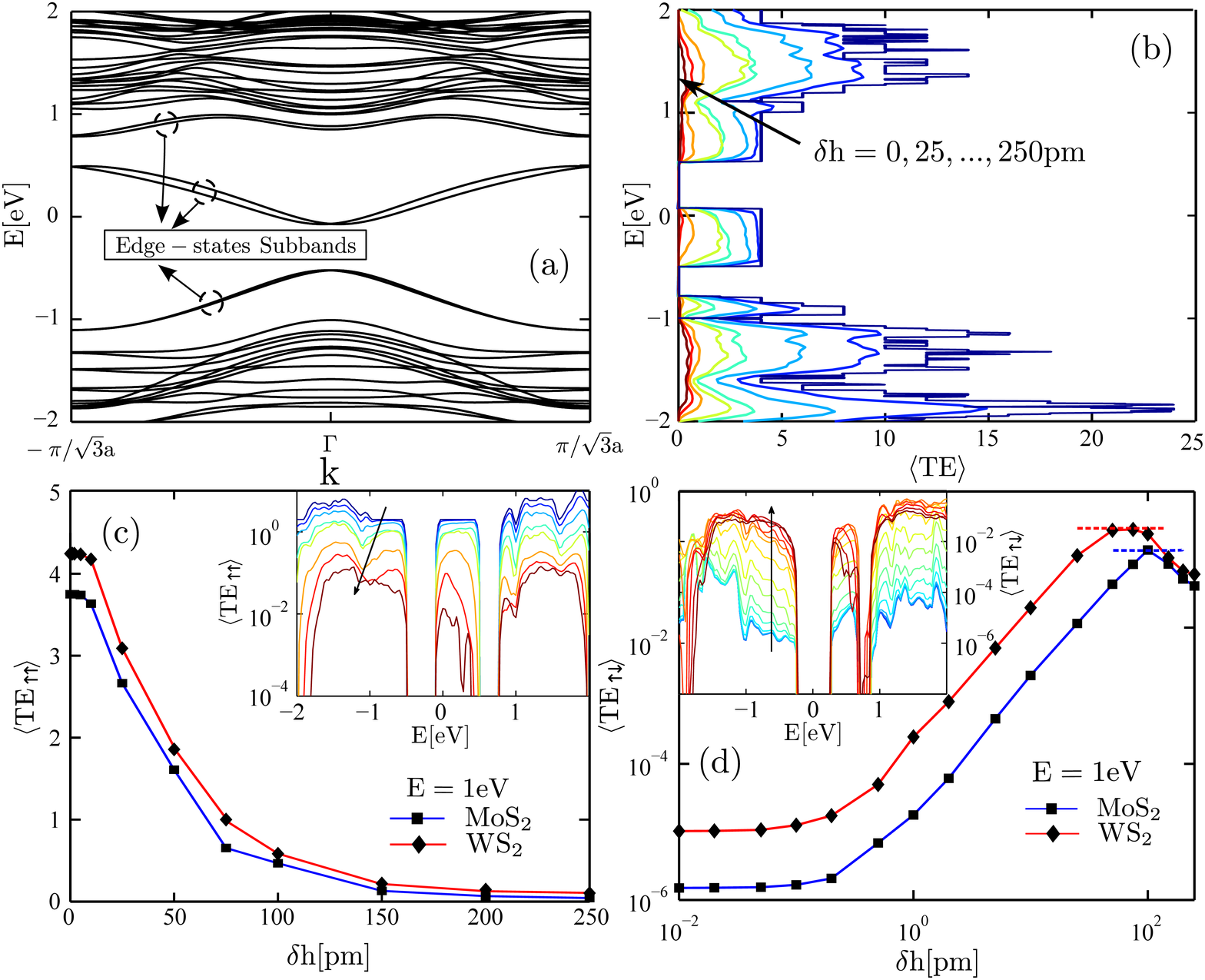}
  \caption{ (a) The band structure of an armchair MoS$_2$ nanoribbon. All the bands are doubly degenerated due to spin. (b) The ensemble average of the total transmission $T$ as a function of energy for different roughness amplitudes. (c) $T_{\uparrow\uparrow}$ as a function of roughness amplitude for MoS$_2$ and WS$_2$. The inset shows $T_{\uparrow\uparrow}$ versus energy for MoS$_2$. (d) $T_{\uparrow\downarrow}$ as a function of $\delta h$ for armchair WS$_2$ and MoS$_2$ nanoribbons. The inset shows $T_{\uparrow\downarrow}$ versus energy for WS$_2$. All results for $nW=15$ and $L=20$~nm.}
  \label{f:MoS2}
\end{figure*}

\begin{figure}[t]
    \includegraphics[width=1.\linewidth]{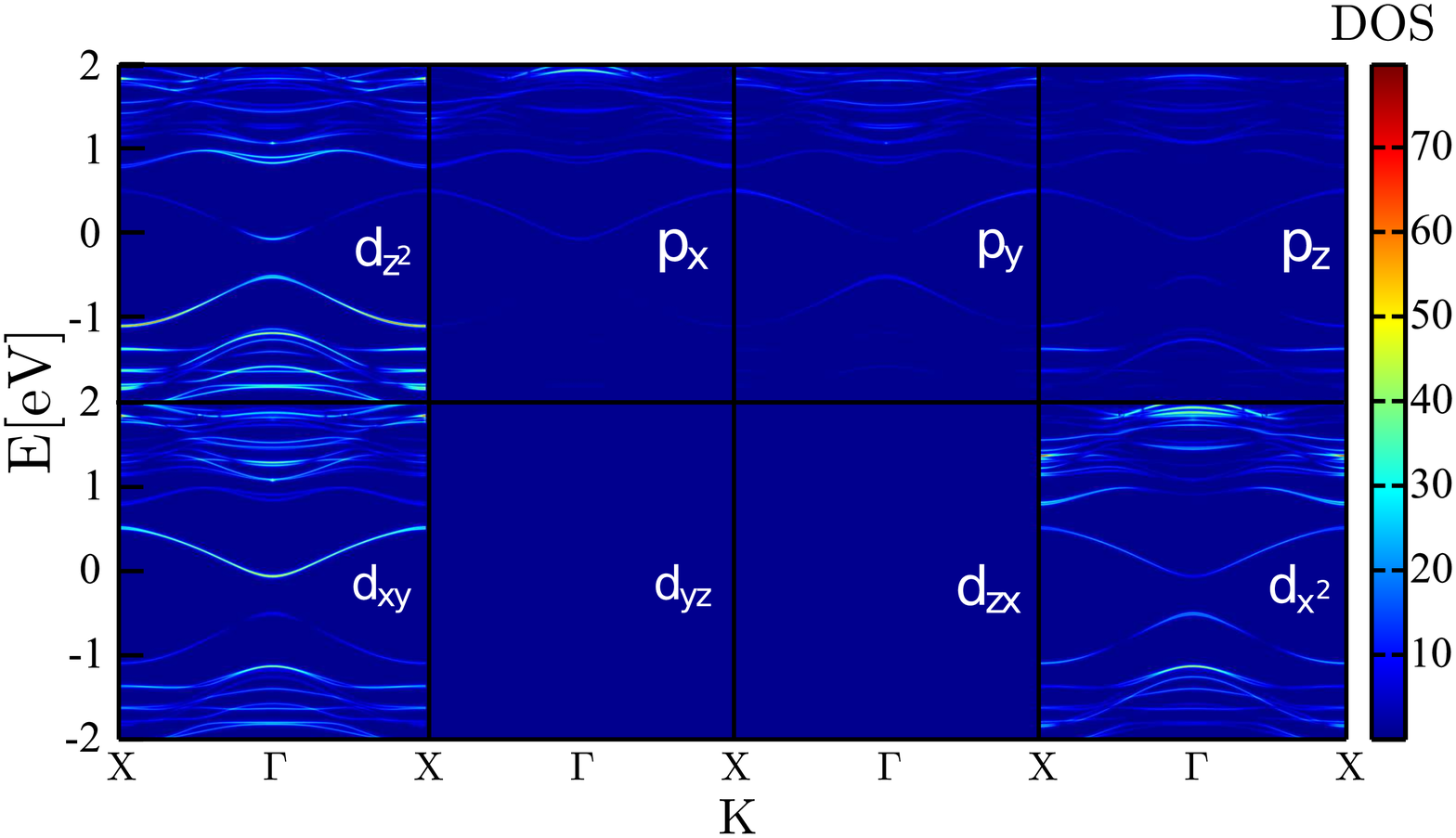}
  \caption{Orbital character of the band structure of the MoS$_2$ nanoribbon shown in Fig. \ref{f:MoS2} (a). Each panel represents  the orbital weight of the corresponding band, where the labels refer to the $d$-character of Mo atom  ($d_{3z^2-r^2}$, $d_{xz}$, $d_{yz}$, $d_{x^2-y^2}$ and $d_{xy}$), and the $p$-character of the chalcogen atom S ($p_x$, $p_y$ and $p_z$). The color scale indicates the corresponding orbital contribution. SOC is not included in this figure.}
  \label{f:Orbital}
\end{figure}

The non-equilibrium Green's function (NEGF) formalism \cite{pourfath14non} is used to study spin transport in armchair MoS$_2$ and WS$_2$ nanoribbons. The channel and the two contacts are assumed to be of the same material (see Fig. \ref{f:Sketch}). For the calculation of the contact self-energies, the surface Green's function of the contacts is iteratively solved, employing a highly convergent scheme \cite{sancho84,sancho85}
\begin{equation}\label{Eq:g}
\underline{g}^{L,R} = \left[E\underline{I}-\underline{H}^{L,R}-\underline{h}_{c}^{L,R} \underline{g}^{L,R}\left(\underline{\tau}_{c}^{L,R}\right)^\dagger\right]^{-1}\ ,
\end{equation}
where $E$ is the energy, $\underline{I}$ is the identity matrix, $\underline{H}^{L,R}$ is the Hamiltonian of the unit cell of the right or left contact in real space representation, $\underline{h}_c^{L,R}$ is the coupling between two neighboring unit cells in the considered contacts, and $\underline{\tau}_{c}^{L,R}$ is the coupling between the channel and the contacts. Underlined quantities stand for matrices that include both spins.  A flat interface is assumed in the contacts, and our calculations include SOC  in the channel as well as in the leads. The retarded $\underline{G}^r$ and advanced $\underline{G}^a$ Green's functions of the device region are then given by:
\begin{equation}
\begin{split}
\underline{G}^r(E)&=[(E+i\delta)\underline{I}-\underline{H}-\underline{\Sigma}^L - \underline{\Sigma}^R]^{-1},\\
\underline{G}^a(E)&=[(E-i\delta)\underline{I}-\underline{H}-\underline{\Sigma}^L - \underline{\Sigma}^R]^{-1},
\end{split}
\end{equation}
where $\delta$ is a phenomenological broadening (10$^{-5}$eV), and $\Sigma_{\sigma}^{L,R}$ is the self-energy of the left and right contacts
 \begin{equation}
\Sigma^{L,R}_{\sigma} = \tau^{L,R}_{\sigma}g^{L,R}_{\sigma}
\left(\tau^{L,R}_{\sigma}\right)^\dagger,
\label{Eq:sigma}
\end{equation}
where $g^{L,R}_{\sigma}$ is the surface Green's function of the contacts, given by Eq. (\ref{Eq:g}). The transmission probability is given by
\begin{equation}\label{Eq:TE}
T(E)={\rm Tr}\left[\underline{\Gamma}^L \underline{G}^r \underline{\Gamma}^R \underline{G}^a\right],
\end{equation}
where $\underline{\Gamma}^{L,R} = i\left(\underline{\Sigma}^{L,R} -\left( \underline{\Sigma}^{L,R}{}\right)^\dagger\right)$  
describes the broadening of the two semi-infinite leads. The spin-resolved transmission probability can be written as:  
\begin{equation} 
\begin{split}
T_{\sigma\sigma'}(E) &= {\rm Tr} \left[ \Gamma^L_{\sigma}G^r_{\sigma\sigma'}\Gamma^R_{\sigma'}G^a_{\sigma'\sigma} \right],
~~~~\sigma,~\sigma'=\uparrow,\downarrow,
\end{split}
\label{Eq:fourTE}
\end{equation}
where $T_{\uparrow\uparrow}(E)$ and $T_{\downarrow\downarrow}(E)$ represent parallel spin transmission, and $T_{\uparrow\downarrow}(E)$ and $T_{\downarrow\uparrow}(E)$ antiparallel spin-flip transmission.

\begin{figure*}[]
    \includegraphics[width=0.8\linewidth]{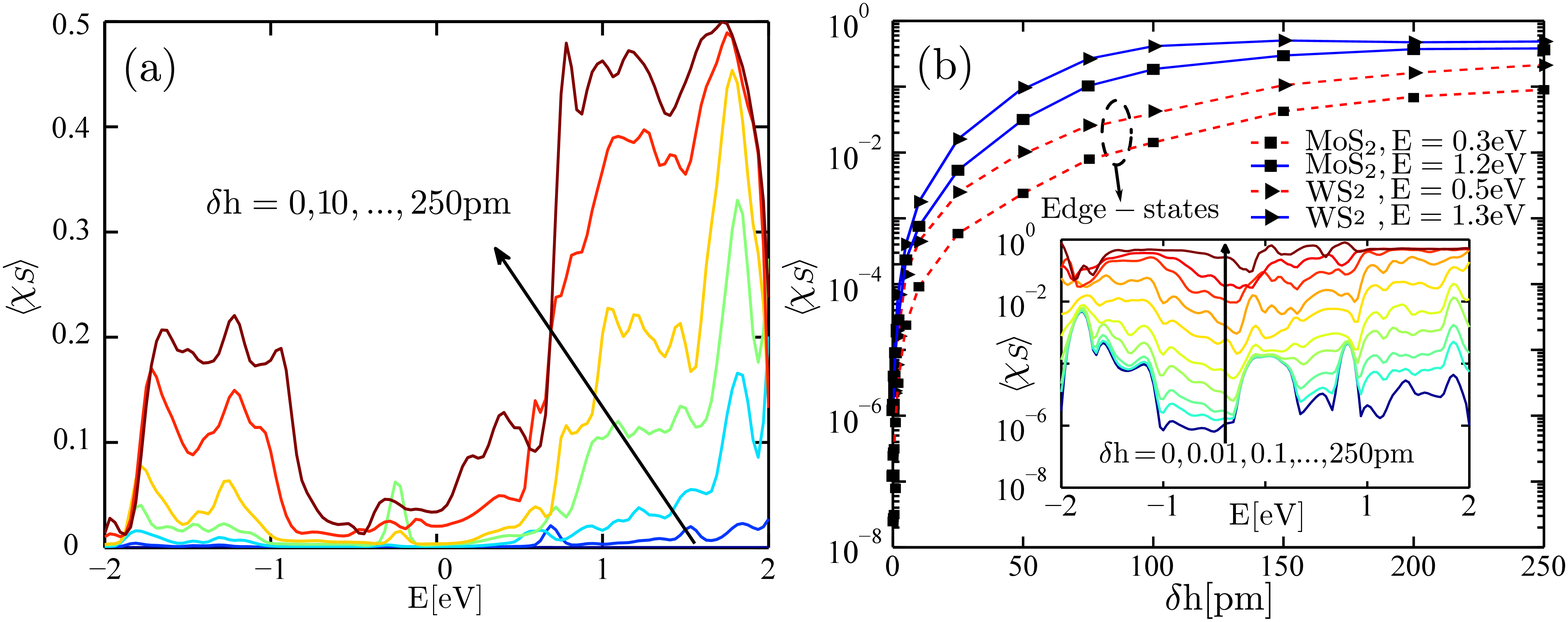} 
  \caption{(a) Spin-flip relative transmission as a function of energy at various roughness amplitudes in armchair MoS$_2$. (b) Spin-flip relative transmission versus $\delta h$ for both MoS$_2$ and WS$_2$ nanoribbons. The inset shows SFRT in logarithmic scale for WS$_2$. All results for $nW=15$ and $L=20$~nm.}
    \label{f:SFRT}
\end{figure*}

\subsection{Effect of lattice modulation}

Different modulations of the lattice have been studied, as bending, sinusoidal ripples and Gaussian corrugation. Here we present results for
the Gaussian corrugation although some comments to the other type of disorder will be made. 
The surface roughness of the substrate is modeled by a Gaussian auto-correlation function:\cite{goodnick85,touski13}
\begin{equation}
 R(x,y)= \mathit{\delta h}^2\exp\left(-\frac{\mathit{x}^2}{\mathit{L_x}^2}-
 \frac{\mathit{y}^2}{\mathit{L_y}^2} \right)\ ,
\end{equation}
where $L_x$ and $L_y$ are the roughness correlation lengths along the $x$ and $y$-directions, respectively, and $\delta h$ is the root mean square of the variation amplitude. We use in most of our calculations $L_x=L_y=40$~nm. Typical roughness parameters for several common substrate materials are reported in Table \ref{tab:rms}. As compared to the well studied case of graphene, we notice that the larger bending rigidity of MoS$_2$ causes smoother surface roughness and longer correlation lengths. Surface roughness modulates atomic positions and orbital directions, thus affecting the hopping parameters.

The effect of lattice deformations due to corrugation are considered within our  Slater-Koster tight-binding model. If we neglect the corrections to the on-site potentials due to lattice deformation, the effect of strain is thus considered  by varying the interatomic bond lengths as a result of the applied strain.
The modified hopping terms in the modulated lattice can be written, at the leading order, as
\begin{eqnarray}\label{Eq:hopping}
V_{i,j;l,m}({\bf r}_{ij})&=&V_{i,j;l,m}({\bf r}_{ij}^0)
\left(1-\beta_{i,j;l,m}
\frac{|{\bf r}_{ij}-{\bf r}_{ij}^0|}{|{\bf r}_{ij}^0|}
\right),
\end{eqnarray}
where $|{\bf r}_{ij}^0|$ is the distance
between two atoms labelled by $i$ and $j$ in the unperturbed lattice,
$|{\bf r}_{ij}|$ the separation in the presence of corrugation,
and $\beta_{i,j;l,m}=-d\ln V_{i,j;l,m}(r)/d\ln(r)|_{r=|{\bf r}_{ij}^0|}$
is the dimensionless bond-resolved local electron-phonon coupling, where $l$ and $m$ are the considered orbitals in atoms $i$ and $j$.
The lattice distances are $|{\bf r}_{ij}^0|=a$ 
for the in-plane $M$-$M$ and $X$-$X$ bonds, where $a$ is the in-plane inter-atomic distance, and $|{\bf r}_{ij}^0|=\sqrt{7/12}~a$
for the $M$-$X$ bond. In the absence of any theoretical and
experimental estimation for the electron-phonon coupling,
we adopt the Wills-Harrison argument~\cite{Harrison:99}, assuming that $\beta_{i,j;l,m}$ depend solely on the total angular momentum of the $l$ and $m$ orbitals, not on their $L^z$ projections.
Namely $V_{i,j;l,m}(r) \propto |{\bf r}|^{-(\ell_l+\ell_m+1)}$, where $\ell_l$ is the absolute value of the angular momentum
of the orbital $l$, and $\ell_m$ is the absolute value of the angular momentum
of the orbital $m$. Following this approach we assume that
$\beta_{i,j;p-p}=3$, $\beta_{i,j;p-d}=4$,
and $\beta_{i,j;d-d}=5$, for the $X$-$X$ $pp$, for $X$-$M$ $pd$, and
for the $M$-$M$ $dd$ hybridizations, respectively. This approximation has been successfully applied to the study of TMDCs in the presence of non-uniform profiles of strain.\cite{Rostami2015b,San-Jose2016} Importantly, this set of parameters matches the direct-to-indirect bandgap transition in MoS$_2$ under $2-3\%$ of biaxial strain as obtained from ab-initio calculations.\cite{Feng:NP12,Wang:ADP14} 

As explained in Sec. \ref{Sec:TB}, we consider here an intrinsic SOC term in the whole Brillouin zone, including
both metal $d$  and chalcogen $p$ orbitals. This term  given by Eq. (\ref{Eq:HSO})
 includes all the spin contributions arising from the crystal potential. The spin-flip terms
of the SOC, as discussed before, are negligible in the flat geometries.
In the corrugated ribbon, the break of the mirror symmetry  produces  non-zero matrix elements between the even and odd
blocks of the Hamiltonian and therefore spin-flip terms of the ${\hat H}_{SO}$ become significative.
In the rest of the paper we present the results for the spin transport properties of corrugated TMDC ribbons obtained by using the above numerical methods. 

\begin{table}[t!]
\caption{The root mean square of surface roughness for TMDCs on various substrates.\cite{yamamoto215self, sercombe2013optical,quereda2014single, azcatl14mos2,yu2013controlled,ji2013epitaxial}} 
\centering 
\begin{tabular}{ccccccc} 
\hline\hline 
SiO$_2$ & ~~~ & Mica & ~~~ & {h-BN} & ~~~ & Al$_2$O$_3$ \\ [0.5ex] 
\hline 
108-420 pm & ~~~ & 77 pm & ~~~ & 63 pm & ~~~ & 140-390 pm \\
\hline 
\end{tabular}
\label{tab:rms}
\end{table}

\section{Results and discussion}
\label{Result} 

\begin{figure}[]
    \includegraphics[width=0.8\linewidth]{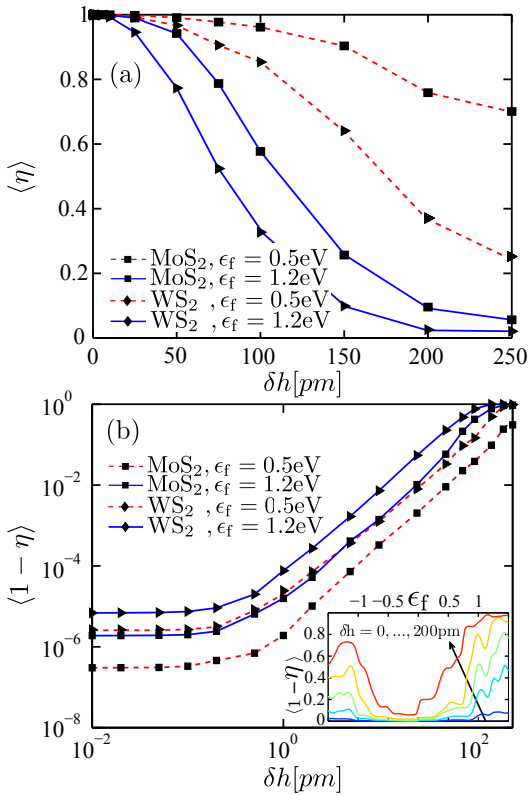}    
  \caption{(a) The normalized spin polarized conductance ($\eta$) as a function of surface roughness amplitude for  MoS$_2$ and  WS$_2$ for two different values of Fermi energy. (b) Logarithmic $1-{\eta}$ versus logarithmic surface roughness amplitude can better describe the behaviour of $\eta$ for low $\delta h$.  The inset show $\eta$ versus Fermi energy at various surface roughness. All results for $L=20$~nm.}
  \label{f:P}
\end{figure}

\begin{figure}[]
    \includegraphics[width=0.8\linewidth]{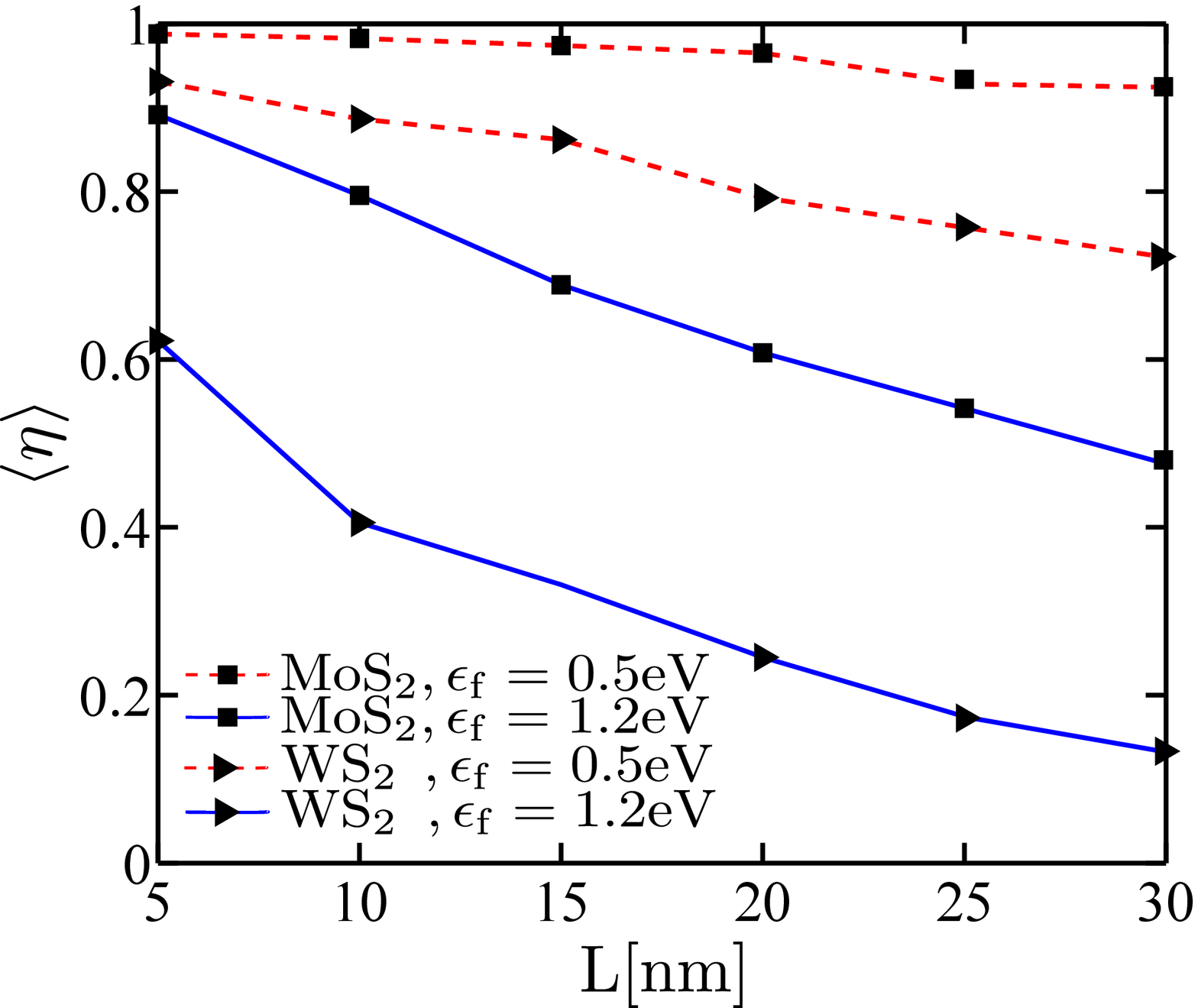} 
  \caption{$\eta$ as a function of the channel length for MoS$_2$ and WS$_2$ nanoribbons. Results for $\delta h=100$~pm, $nW=15$, $L_x=L_y=50$~nm.}
  \label{f:PL}
\end{figure}

In graphene, it is known that surface roughness mixes $\pi$ with $\sigma$ bonds, what enhances spin-orbit interaction.\cite{huertas06,Chico09,lopez11} In the following we will show which is the effect of sample corrugations on the charge and spin transport properties of TMDCs.
The band structure of a MoS$_2$ armchair nanoribbon is shown in Fig. \ref{f:MoS2}(a).  In agreement with density functional theory (DFT) calculations,\cite{li08mos2,lopez2015electronic} our tight-binding model for the ribbon with this termination shows a semiconducting behaviour, with the appearance of edge states inside the gap, that would be absent if periodic boundary conditions are considered. These edge-state subbands  are marked in Fig. \ref{f:MoS2}(a). It is important to notice that the energy bands are spin degenerated and they are split away from the time reversal invariant $\Gamma$ point of the Brillouin zone due to the effect of SOC.\cite{Rostami2016}  The orbital character of each band is shown in Fig. \ref{f:Orbital}. It is interesting to notice that, due to the band folding characteristic of a nanoribbon, the bands at the $\Gamma$ point present an important contribution from $d_{3z^2-r^2}$, $d_{x^2-y^2}$ and $d_{xy}$ for both, valence and conduction states.  The ensemble average of the total transmission probability for this system is calculated, by using Eq. (\ref{Eq:TE}), for several amplitudes of surface roughness with $L_x=L_y=40$~nm, and the results are shown in Fig. \ref{f:MoS2}(b). In the absence of surface corrugation  ($\delta h=0$) the transmission at a given energy $E$ is equal to the number of available subbands at that particular energy. For example, around $E=0$ we observe that $\langle TE \rangle \approx 4$ for $\delta h=0$ (blue line in Fig. \ref{f:MoS2}(b)), which corresponds to the contribution of the two pairs of subbands observed in Fig. \ref{f:MoS2}(a), which are doubly degenerated due to spin. Therefore we observe that the transmission probability for flat nanoribbons is almost unaffected by the spin-orbit interaction.

Realistic samples, however, present surface roughness that modulates the distance and overlap between atomic orbitals. This results in local variations of the hopping parameters and break the surface mirror symmetry. As a consequence, the total transmission decreases with the amplitude of the corrugations. This evolution is shown by the arrow in Fig. \ref{f:MoS2}(b), that shows how the average total transmission $\langle TE\rangle$ decreases with the corrugation amplitude $\delta h$.

 In order to investigate the role of surface roughness on spin transport, the spin-resolved transmission have been calculated: $T_{\uparrow\uparrow}$ and $T_{\uparrow\downarrow}$ are shown in Fig. \ref{f:MoS2}(c)-(d) as a function of $\delta h$ for MoS$_2$ and WS$_2$. It can be observed that $T_{\uparrow\uparrow}$ (panel (c)) and $T_{\uparrow\downarrow}$ (panel (d)) present opposite trend with the amplitude of surface corrugation $\delta h$. As expected the spin-conserved transmission $T_{\uparrow\uparrow}$ (like the the total transmission $\langle TE\rangle$, panel (b)) decreases with $\delta h$. This behaviour is due to the enhancement of the spin-flip processes in Eq. (\ref{Eq:TE}) induced by the variation of the hoppings associated to the sample corrugation.
 In fact $T_{\uparrow\downarrow}$,  an indication of spin-flip rate, increases with $\delta h$, as shown in Fig. \ref{f:MoS2}(d). Both quantities, $T_{\uparrow\uparrow}$ and $T_{\uparrow\downarrow}$, are larger for WS$_2$ (diamonds) than for MoS$_2$ (squares). In particular, $T_{\uparrow\downarrow}$ for WS$_2$ is approximately an order of magnitude larger than that of MoS$_2$ (Fig. \ref{f:MoS2}(d)). This is expected since WS$_2$ has a larger intrinsic spin-orbit coupling than MoS$_2$ (see Table \ref{Tab:Parameters}), which results in a stronger Rashba-like spin-orbit interaction induced by surface roughness, with the corresponding enhancement of spin-flip processes.

The inset of Fig. \ref{f:MoS2}(d) clearly shows that $T_{\uparrow\downarrow}$ increases exponentially with roughness amplitude of up to four order of magnitude over a variation of $\delta h$ from $\sim 10^{-2}$~pm to $\sim10^2$~pm.
Interestingly, our calculations also suggest that $T_{\uparrow\downarrow}$ reaches a maximum  for a given value of $\delta h$, and then it starts to decrease. This threshold is about $\approx 75$~pm for MoS$_2$ and $\approx\mathrm{100 }$~pm for WS$_2$ (see Fig. \ref{f:MoS2}(d)). As explained before two scattering mechanisms affect spin transport: surface corrugation and spin-orbit interaction, which is enhanced with surface roughness. It is also interesting to notice that the spin-flip scattering rate is similar for electron and hole sectors. Since we are dealing with ribbon geometry both,  the edge states and  the folding of the Brillouin zone, play an important  role. In particular, it is very important to notice that the bands (Fig. 2a) are spin degenerate (for both, electron and hole sectors) for the armchair nanoribbons considered here. This is completely different to the case of bulk single layer or zigzag nanoribbons,\cite{Rostami2016} where spin-valley coupling is more robust for valence band states, since valence band edges at K and K' valleys correspond to opposite spin, and they are well separated in energy from the other subband (the separation is $\sim$150 meV for MoS$_2$ and $\sim$400 eV for WS$_2$). The armchair termination is a line of dimers with atoms of the two sublattices, therefore the edge states present valley mixing, as it happens in graphene armchair ribbons (see e.g. Refs. \onlinecite{Brey06,lopez11,Yuan_PRB_2013}). Furthermore, the orbital contribution for the low energy states of both, valence and conduction bands, are rather similar, as it can be seen in Fig. \ref{f:Orbital}. This is due to the band folding that happens for a finite ribbon, with the result that, at the $\Gamma$ point of the ribbon BZ, there are contributions from 'bulk states' at $\Gamma$ and K points. The consequence of this band reconstruction is that, due to the spin degeneracy of the bands and the similar orbital character for electron and hole sectors, the spin-flip scattering probability is of the same order in the two cases.

We have considered other kinds of corrugations like periodic sinusoidal rippling of the sample. We have observed that this kind of corrugation, which can be induced in the laboratory by using elastomeric substrates,\cite{Quereda2016} leads to qualitatively similar effects in the spin-flip transmission $T_{\uparrow\downarrow}$ (not shown here) but of much weaker magnitude as compared to random Gaussian modulation. We have also checked that bending of the ribbon leads to reduction of $T_{\uparrow\downarrow}$ as the curvature radius increases. For this case, the polarised transmission is also much smaller than the obtained with the Gaussian corrugation.

 It is interesting to calculate the spin-flip relative transmission (SFRT) $\chi_S = T_{\uparrow\downarrow}/(T_{\uparrow\uparrow}+T_{\uparrow\downarrow})$, which is a measure of the efficiency of spin control.\cite{michetti10} As expected, the ensemble average of $\chi_S$ as a function of energy (Fig. \ref{f:SFRT}(a)) increases with surface roughness amplitude. We next compare $\chi_S$ for edge- and bulk-states, as indicated in Fig. \ref{f:SFRT}(b). The amplitude of the edge-state wave functions across the width of the armchair ribbon is originated mainly from $d$-orbitals of the metal (Mo or W) and it is localised at the border atoms, decreasing with the distance to the edge. Our results show that bulk-states are strongly affected by surface roughness, whereas edge-states are more robust against corrugations which results in a lower $\chi_S$.  This is expected since the spatial localization of edge states protect them partially from surface corrugation effects. The results for transport from purely bulk states are shown in Appendix \ref{App:Periodic}, where we show simulations with periodic boundary conditions.   
At high energies and large $\delta h$, $\chi_S$ reaches 0.5 that implies a complete loss of spin-information during transmission. This clearly suggest that substrates with rough surfaces, such as the most commonly used SiO$_2$, are not appropriate for spintronic applications based on TMDC materials (see Table \ref{tab:rms}).

The  spin-flip ratio ($\eta$) is another important figure of merit for spintronic devices, defined as
\begin{equation}
\eta=\frac{{\cal G}_{sc}-{\cal G}_{sf}}{{\cal G}_{sc}+{\cal G}_{sf}},
\end{equation}
where ${\cal G}_{sc}={\cal G}_{\uparrow\uparrow}+{\cal G}_{\downarrow\downarrow}$ and ${\cal G}_{sf}={\cal G}_{\uparrow\downarrow}+{\cal G}_{\downarrow\uparrow}$, are the spin-conserving and spin-flip conductances respectively.\cite{chico15} The conductance in the linear regime is given by\cite{Ryndyk_Book09}
\begin{equation}
{\cal G}_{\sigma\sigma'}={\cal G}_0\int_{-\infty}^{+\infty} dE\left(-\frac{\partial f(E-\epsilon_f)}{\partial E}\right) T_{\sigma\sigma'}(E) \ ,
\end{equation}
where ${\cal G}_0=e^2/h$, $f(E)$ is the Fermi-Dirac distribution and $T_{\sigma\sigma'}(E)$ is the transmission, Eq. (\ref{Eq:fourTE}). The position of the Fermi level affects the conductance. 
The results of our calculations are plotted, as a function of roughness amplitudes $\delta h$, in Fig. \ref{f:P} for $\epsilon_f=0.5$~eV, coinciding with an edge state band, and for $\epsilon_f=1.2$~eV, which crosses the bulk bands.  The reduction of $\eta$ with $\delta h$ suggests again that  the spin-flip rate is increased by the surface corrugation. 
Smaller $\eta$ is observed for higher values of the Fermi energy. This is due to the smaller effect of corrugations on edge states and to the larger density of states at high energies, which cause more spin-flip processes. A similar effect has been observed in graphene.\cite{touski13}

Finally in Fig. \ref{f:PL} we show the dependence of $\eta$ with the channel length, for the the same values of the Fermi energy. We observe that the decay of $\eta$ with the channel length is more pronounced for $\epsilon_f=1.2$~eV (crossing bulk states) than for $\epsilon_f=0.5$~eV (crossing edge states). This can be understood again from the fact that spin transport in the second case occurs mainly through the edges, for which we have seen that the effect of corrugations is small. However, when the Fermi level crosses the bulk states, strongly affected by corrugations, $\eta$ decreases faster with the length of the channel.

\section{conclusion}\label{Sec:Conclusions}
In summary, we have performed a systematic theoretical study on spin transport in MoS$_2$ and WS$_2$ armchair nanoribbons in the presence of surface roughness.  
In  the perfectly flat ribbons, the spin-flip terms contribution are negligiblie. Nonetheless, when surface roughness is present,  surface mirror symmetry or $z$-axis symmetry is broken  generating an additional  Rashba-like contribution to the spin-orbit coupling. The strength of this coupling is proportional to the atomic SOC and increases with the corrugation amplitude. Deformation of the surface by ripples, bending or corrugation, modulates the atomic positions thus changing the atomic interactions and orbital hybridisation. The results indicate that sample corrugations significantly enhance the  spin-flip rate. For the same surface roughness,  the  spin-flip rate is larger in WS$_2$ than in MoS$_2$ due to the stronger intrinsic SOC of  heavier W atoms. Our results indicate that the spin information can be completely suppressed in TMDCs-based channel  with armchair termination supported on rough substrates, such as SiO$_2$. Therefore, the use of flat substrates or the application of techniques to stretch the MoS$_2$ or WS$_2$ samples, avoiding undesirable corrugations, can improve the performance of TMDCs based spintronics devices.



 \begin{acknowledgments}

R.R. acknowledges financial support from MINECO (Spain) through grant FIS2014-58445- JIN.
M.P.L.S. acknowledges financial support
by the Spanish MINECO grant
FIS2014-57432-P, the European Union structural funds
and the Comunidad de Madrid MAD2D-CM Program (S2013/MIT-3007).

\end{acknowledgments}

\begin{appendix}

\section{Simulations with periodic boundary conditions}\label{App:Periodic}

\begin{figure}[]
    \includegraphics[width=0.8\linewidth]{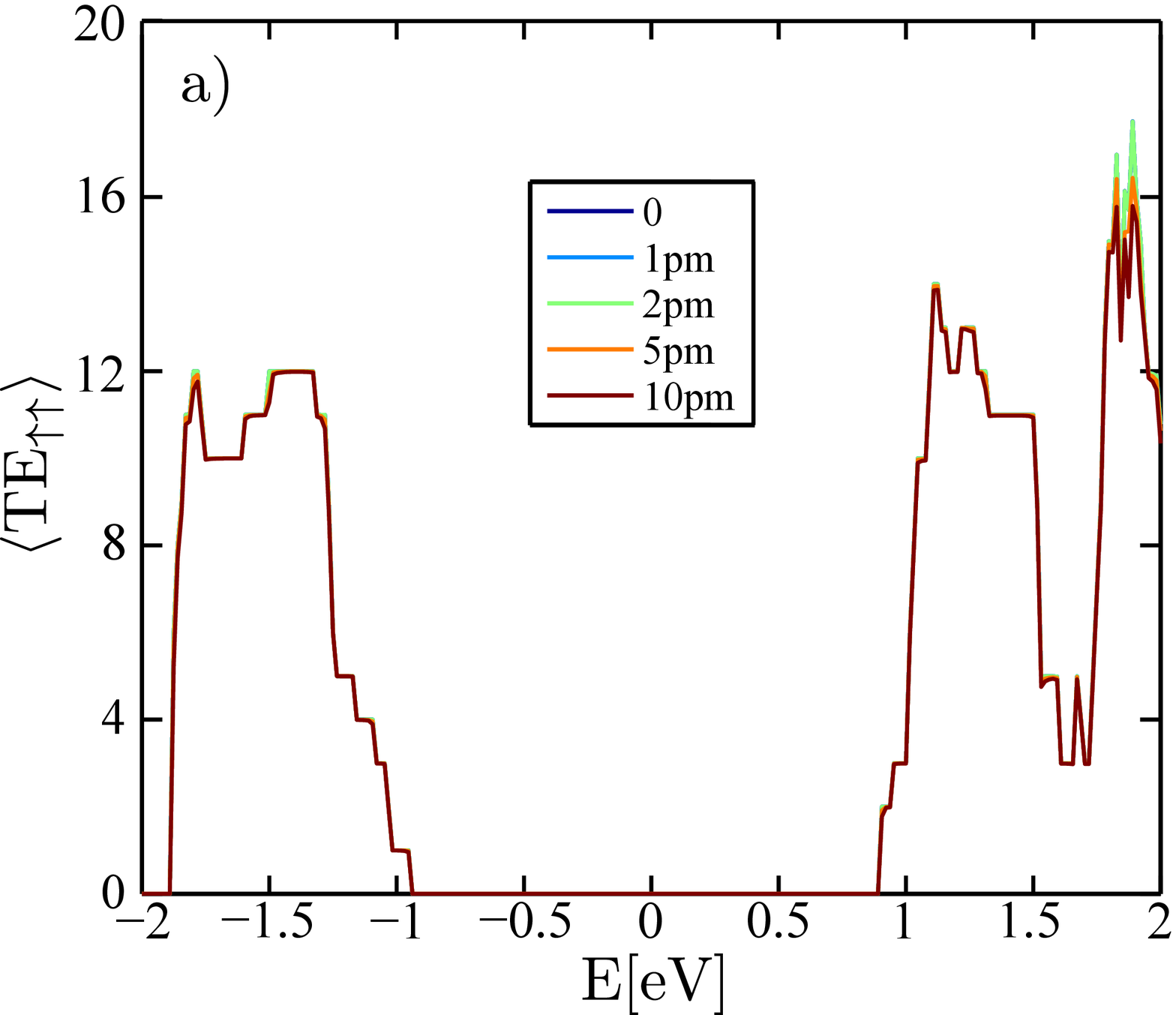} 
    \includegraphics[width=0.8\linewidth]{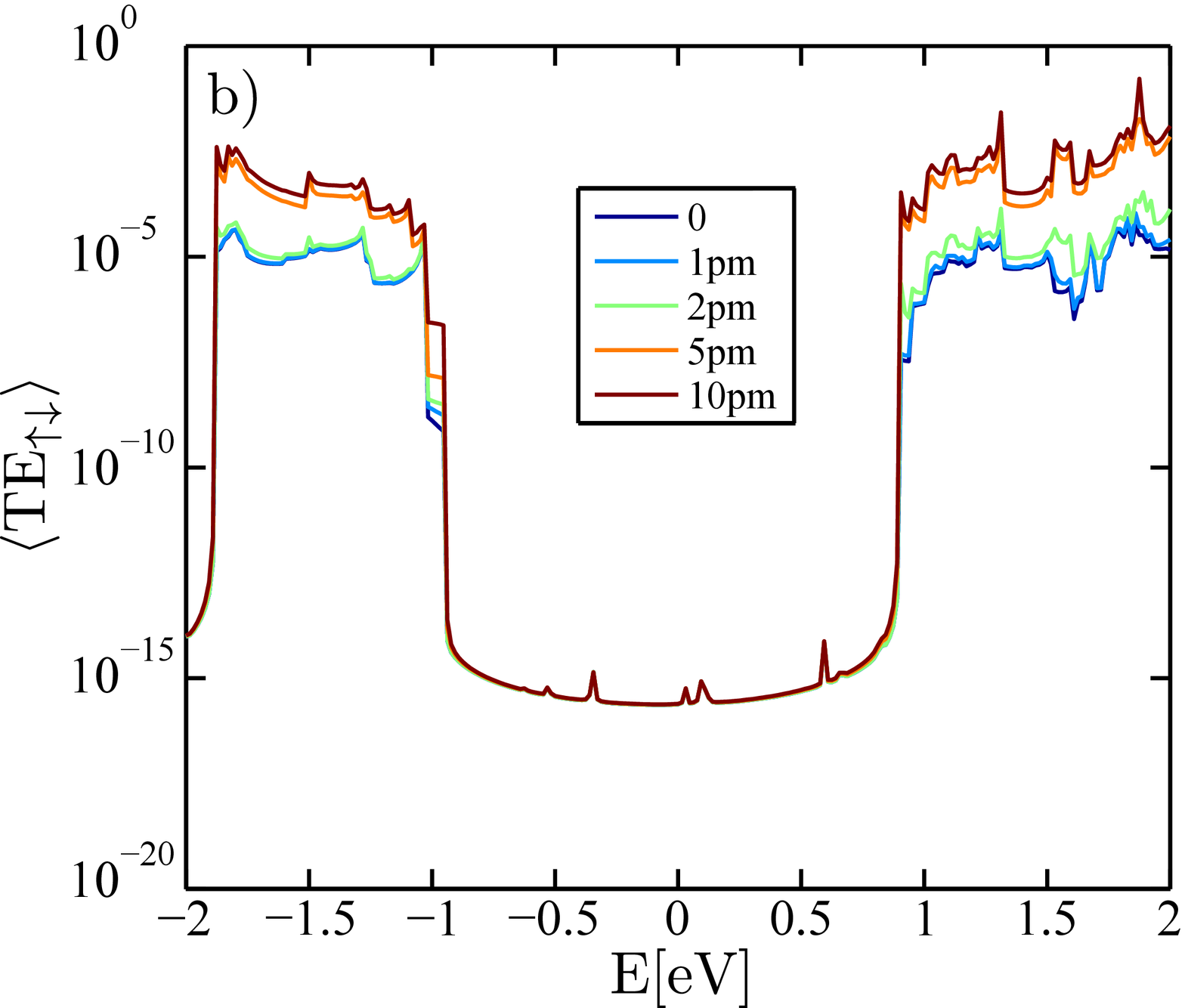} 
  \caption{(a) $T_{\uparrow\uparrow}$ of MoS$_2$ as a function of energy, using periodic boundary conditions, for different values of $\delta h$. (b) Same as (a) but for $T_{\uparrow\downarrow}$. All results for $nW=15$ and $L=20$~nm.}
  \label{f:PeriodicBoundary}
\end{figure}

In order to identify the contribution to transport from purely bulk states, in this appendix we present results of calculations done with periodic boundary conditions. The results are shown in Fig. \ref{f:PeriodicBoundary}. Here the amplitud of the corrugations ($\sim$10 pm) is much smaller than the used for open boundary conditions (up to $\sim$ 250 pm). This is due to a technical difficulty to obtain the same corrugation in the two edges of the nanoribbon to be connected when periodic boundary conditions are considered. Even for such small corrugations, we observe that $TE_{\uparrow\downarrow}$ due to purely bulk states (there is no edge states present in this calculaton) increases in more than two order of magnitude from a flat nanoribbon to one with corrugations of $\sim 10$ pm amplitude. 

\end{appendix}


\end{document}